# Ultrasensitive nanoelectromechanical mass detection

K.L. Ekinci,[a)] X.M.H. Huang, and M.L. Roukes [b)]

*Condensed Matter Physics, 114-36, California Institute of Technology, Pasadena, CA 91125*

We describe the application of nanoelectromechanical systems (NEMS) to ultrasensitive mass detection. In these experiments, a modulated flux of atoms was adsorbed upon the surface of a 32.8 MHz NEMS resonator within an ultrahigh vacuum environment. The mass-induced resonance frequency shifts by these adsorbates were then measured to ascertain a mass sensitivity of $2.53 \times 10^{-18}$ g. In these initial measurements, this sensitivity is limited by the noise in the NEMS displacement transducer; the ultimate, limits of the technique are set by fundamental phase noise processes. Our results and analysis indicate that mass sensing of individual molecules will be realizable with optimized NEMS devices.

Nanoelectromechanical[1] systems (NEMS) are emerging as strong candidates for a host of important applications in semiconductor-based technology and fundamental science.[1] The minuscule active masses of NEMS, in particular, render them extremely sensitive to added mass — a crucial attribute for a wide range of sensing applications.

Resonant mass sensors with high mass sensitivities have been employed in many diverse fields of science and technology. Among the most sensitive are those based on the acoustic vibratory modes of crystals,[2,3] thin films[4] and micron-sized cantilevers.[5,6,7,8] In all of these, the vibratory mass of the resonator, its resonance frequency, and quality factor ($Q$) are central in establishing its mass sensitivity. In this Letter, we demonstrate attogram-scale inertial mass sensing using high-frequency NEMS, and discuss how even greater sensitivity will be obtainable with such devices. This provides a concrete initial demonstration of the potential that nanoscale mechanical devices offer for sensing and ultimately weighing individual molecules.

These initial experiments were carried out in an ultra high vacuum (UHV) environment within the apparatus depicted in Figure 1(a). This system allows the operation of a NEMS resonator (transduced magnetomotively[9]) while a pulsed, weak flux of Au atoms is directed upon it. The Au atoms are generated by a thermal evaporation source and travel ballistically towards the NEMS within the apparatus. The mass flux, $F$, of the evaporator is measured by a calibrated quartz crystal monitor and modulated by a shutter; both are in the vicinity of the evaporator. The resonator temperature is regulated at $T \approx 17$ K, both to ensure unity adsorbate sticking probability[10] and to allow careful monitoring of the resonator temperature fluctuations (see Fig. 3). Then, with knowledge of the exposed NEMS surface area, $S$, (determined from careful scanning electron microscopy measurements) we can determine the *exact mass*[11] of the adsorbed Au atoms on the NEMS as

$$\Delta m(t) \approx \int_0^t SF(r_{QCM}/r_{NEMS})^2 dt.$$

In this system, the geometric factor, $(r_{QCM}/r_{NEMS})^2 \approx 5 \times 10^{-3}$.

We employed nanomechanical doubly-clamped SiC beam resonators such as the ones shown in Figure 1(b) as the sensor elements in these experiments. The beams are embedded within a radio frequency (RF) bridge configuration, creating a unique two-port device.[12] The fundamental in-plane flexural resonance frequency of one of the resonators in the bridge was tracked continuously by the phase-locked loop (PLL) circuit shown schematically in Figure 1(c). The single suspended beam structure of Fig. 1 (b), labeled T on the right, enables four-wire resistance measurements, which provide extremely sensitive monitoring of the local temperature of the suspended devices (cf. Fig. 3)In Figure 2, we display the temporal evolution of the fundamental-mode resonant frequency of the doubly-clamped beam resonator, as it is exposed to a ballistic flux of Au atoms. Adsorption commences when the shutter (*see* Fig. 1(a)) is opened during specific time intervals — changing the effective resonator mass, $M_{eff}$. The mass responsivity, $\Re = \partial \omega_0 / \partial M_{eff}$ was deduced to be $\Re/2\pi \approx 2.56 \times 10^{18}$ Hz/g from a linear fit to the data points in the steps of the upper and lower plots, *i.e.* $\Delta\omega_0(t)/2\pi$ vs. $\Delta m(t)$. The noise floor of the measurement was determined from the regions of constant frequency when the shutter was closed. For this experiment, a measurement bandwidth of $\Delta f=3$ kHz ($\tau \sim 2$ms) yielded a frequency noise floor, $\delta\omega_0/2\pi \approx 6.51$ Hz — corresponding to a minimum detectable mass, $\delta M = \Re^{-1}\delta\omega_0 \approx 2.53 \times 10^{-18}$ g (ref.13). In units of the atomic mass of Au, $m_{Au}$, $\delta M \approx 7400 m_{Au}$.

We have taken special precautions in these measurements to minimize the thermal frequency fluctuations and drifts, given that the NEMS resonator will be exposed to both hot incoming atoms and blackbody radiation from the thermal source. This is of special concern here since the thermal resistance — and hence the thermalization rate — between a suspended NEMS device and its environment can be exceptionally large.[14] In Figure 3, we show four-probe electrical measurements of the resistance fluctuations, $\delta R/R$, of

---

[a)] Present address: Aerospace & Mechanical Engineering Dept., Boston University, Boston, MA 02215 Electronic mail: ekinci@bu.edu
[b)] Electronic Mail: roukes@caltech.edu

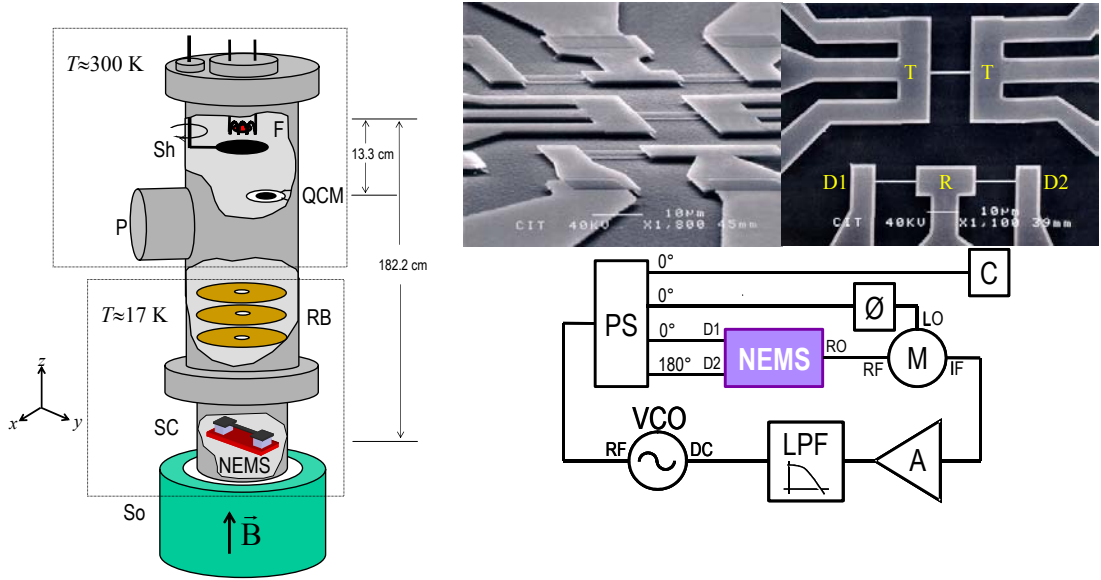

**Figure 1. (a)** Variable temperature, UHV microwave cryostat for mass sensitivity measurements. The sample chamber (**SC**) is inserted into the bore of a 6T superconducting solenoid (**So**) in liquid helium. The radiation baffles (**RB**) establish a line of sight along the *z*-axis from a room temperature thermal-evaporation source (**TS**) to the bottom of the cryostat. The NEMS resonators are carefully placed in this line-of-sight, some $r_{NEMS}$=182.2 cm away from the thermal-evaporation source. A calibrated quartz crystal monitor (**QCM**) at a distance of $r_{QCM}$=13.3 cm and a room temperature shutter (**Sh**) are employed to determine and modulate the atom flux, respectively. **(b)** Scanning electron micrographs of nanomechanical doubly-clamped beam sensor elements. The beams are made out of SiC with top surface metallization layers of 80 nm of Al. The beams are configured in a radio frequency (RF) bridge with corresponding actuation (D1 and D2) and detection (R) ports as shown. The central suspended structure attached to 3 contact pads on each side, labeled T, is for monitoring the local temperature. **(c)** Conceptual diagram of the phase-locked loop (PLL) NEMS readout. The principal components are: voltage controlled radio frequency (RF) oscillator (**VCO**); 4-port power splitter (**PS**) (with three 0° and one 180° output ports); **NEMS** mass sensor with RF bridge readout; mixer (**M**); phase shifter (**Ø**); variable gain amplifier (**A**); low pass filter (**LPF**); frequency counter (**C**).

a metallized suspended SiC beam identical in size to NEMS mass sensors (see Fig. 1(b)). In this measurement, baseline $\delta R/R$ was established when the shutter was in the closed position, *i.e* for $t < 50$ s. The device was exposed to Au atom fluxes comparable to those employed in the mass sensitivity experiments, when the shutter was opened at $t \geq 50$ s. $\delta R/R$ of the metal electrode is converted to temperature fluctuations using measured temperature dependence of the four-terminal resistance. $\partial R/\partial T \approx 0.05$ Ω/K in the vicinity of $T\approx 17$ K (Fig. 3 left inset).[15] Separate measurements of the temperature dependence of the resonance frequency near $T\approx 17$ K yielded $\frac{1}{2\pi}\frac{\partial \omega_0}{\partial T} \approx -700$ Hz/K (Fig. 3 right inset). One can then estimate the thermally induced frequency fluctuations as $\frac{\delta \omega_0}{2\pi} = \frac{1}{2\pi}\frac{\partial \omega_0}{\partial T}\left[\frac{\partial R}{\partial T}\right]^{-1}\delta R \leq 5$ Hz. The frequency fluctuations thus deduced are comparable to our noise floor for zero flux — establishing that no significant thermal effects are observable for the very low fluxes employed in these experiments.

We now turn to a noise analysis of our measurements. In general, resonant mass sensing is performed by carefully determining the resonance frequency of the resonator and then, by looking for a frequency shift in the *steady state* due to the added mass. Hence, to determine the *mass sensitivity*, $\delta M$, one needs to consider the noise floor for frequency measurements, $\delta \omega_0$, since

$$\delta M \approx \left|\frac{\partial M_{eff}}{\partial \omega_0}\right|\delta \omega_0 = |\Re|^{-1}\delta \omega_0. \qquad (1)$$

For the fundamental mode doubly-clamped beam sensors, the effective mass is a fraction of the total resonator mass, $M_{eff} = 0.735 M_{tot} = 0.735 \rho L t w$. Here, $L \times t \times w$ are the beam's dimensions, and $\rho$ its mass density. In the limit $\delta M \ll M_{eff}$, the resonator's characteristics — in particular, its compliance and $Q$ — will be relatively unaffected by mass accretion. In this regime, $|\Re| = \omega_0 / 2 M_{eff}$, and $\delta M \approx 2(M_{eff}/\omega_0)\delta \omega_0$. An estimate for $\delta \omega_0$ can be obtained by integrating the spectral density of the frequency fluctuations, $S_\omega(\omega)$, over the effective measurement bandwidth[16] $\Delta f$:

$$\delta \omega_0 \approx \left[\int_{\omega_0 - \pi \Delta f}^{\omega_0 + \pi \Delta f} S_\omega(\omega) d\omega\right]^{1/2}. \qquad (2)$$

Elsewhere[17] we have carefully analyzed the fundamental physical limits to inertial mass sensing imposed by the frequency-fluctuation noise $\delta \omega_0$. This analysis leads to the conclusion that in these initial experiments, $\delta\omega_0$ and hence, $\delta M$ are both limited by the measurement electronics, *i.e.* the transducer, and readout circuitry. In other words, the spectral density of the voltage noise, $S_V(\omega)$, at the output will determine the measured frequency noise as $S_\omega(\omega) = S_V(\omega)/(\partial V/\partial \omega)^2$ (ref.18) .Here, $\partial V/\partial \omega$ is the rate of change of the transducer output in the vicinity of the resonance frequency. We can crudely approximate this as $\partial V/\partial \omega \approx V_{max}/\Delta\omega = QV_{max}/\omega_0$. Here, $V_{max}$ is the maximal response of the transducer (on resonance) and depends upon the drive level. To maximize the signal-to-noise ratio, one applies the largest rms drive level, $\langle x_{max}\rangle$, consistent with producing predominantly linear response. This, combined with the readout transducer responsivity, $R_T = \partial V/\partial x$ (with units of V/m) leads to the maximal value $(\partial V/\partial \omega)_{max} \approx QR_T\langle x_{max}\rangle/\omega_0$. Upon expressing $S_\omega(\omega)$ in terms of $S_V(\omega)$, and integrating Eq. (2), $\delta\omega_0$, limited by the readout process, is

$$\delta\omega_0 \approx \frac{\omega_0}{Q}\frac{(S_V 2\pi\Delta f)^{1/2}}{R_T\langle x_{max}\rangle} . \quad (3)$$

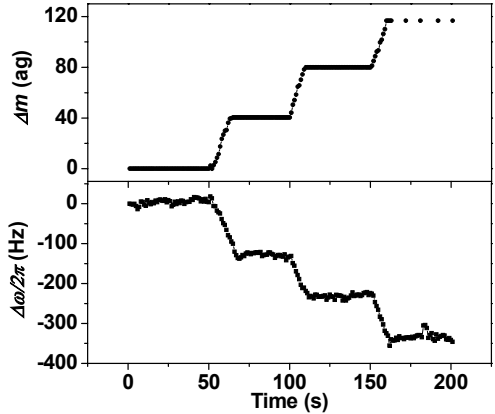

**Figure 2.** Frequency shifts, $\Delta\omega/2\pi$, (bottom) induced by sequential 40 attogram (ag) (1 ag=$10^{-18}$ g) gold atom adsorption upon a 14.2 μm x 670 nm x 259 nm silicon carbide doubly-clamped beam resonator. The (initial) fundamental frequency is $\omega_0/2\pi \approx 32.8$ MHz. The accreted mass of gold atoms, $\Delta m$, in the upper plot is measured by a separate quartz crystal detector. The rms frequency fluctuations of the system (*i.e.* the noise level in the lower trace) correspond to a mass sensitivity of 2.5 ag for the 2 ms averaging time employed.

We have made a simplifying assumption that $S_V(\omega)$ is independent of $\omega$, *i.e.* white, in the measurement band. Eq. (3) depends inversely upon the ratio of the transducer's maximum response to its noise floor (at the output), $R_T\langle x_{max}\rangle/(S_V 2\pi\Delta f)^{1/2}$. Since it is the square of this ratio that defines the (power) dynamic range (DR) of the coupled mechanical resonator/transducer system, we can write

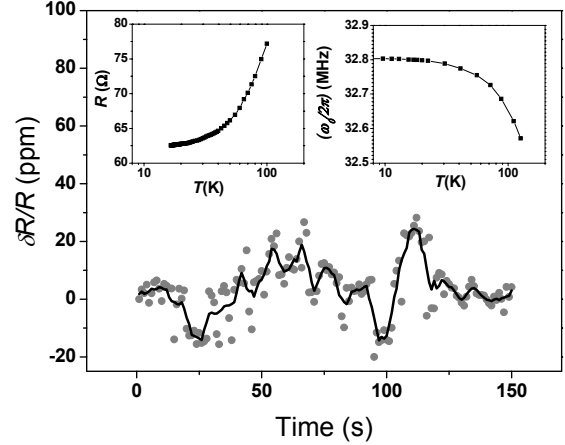

**Figure 3.** The four-terminal resistance fluctuations, $\delta R/R$, of the metal electrode on a suspended SiC beam. $R\approx 62.5$ Ω at $T\approx 17$ K. $\delta R/R$ is converted to local temperature fluctuations using measured values of temperature dependence of the electrode resistance (left inset). The local temperature fluctuations are then converted to thermally-induced frequency fluctuations using the temperature dependence of the resonator frequency (right inset). Here, $\omega_0/2\pi \approx 32.8$ MHz at $T\approx 17$ K.

$\delta\omega_0 \approx (\omega_0/Q)10^{-(DR/20)}$ yielding the simple and compelling expression

$$\delta M \approx 2(M_{eff}/Q)10^{-(DR/20)} . \quad (4)$$

Eq. (4) allows estimation of the mass sensitivity attainable with our 32.8 MHz device with the following (separately-measured) experimental parameters: $Q\approx 3000$, $DR\approx 60$dB for a $\Delta f\approx 3$ kHz bandwidth, and $M_{tot}\approx 9.9\times 10^{-12}$ g. This leads to the approximate result that $\delta M\sim 10^{-18}$ g, quite close to what we experimentally attain. Table 1 summarizes mass responsivity measurements obtained from experiments employing other NEMS devices. The small inconsistencies between the predicted and experimentally-measured inverse responsivities reflect the extreme mass sensitivity of NEMS. The presence of surface adsorbates at sub-monolayer coverage is enough to change perceptibly the experimental device parameters.

Projections based on Eq. (4), and its confirmation as provided in our experiments, make clear that NEMS mass sensing can provide significant advances in chemical and biological sensing and mass spectrometry. With the attainment of NEMS operating at microwave frequencies[19] prospects for detection of individual, electrically-neutral macromolecules with single-Dalton sensitivity become feasible. This is a regime where the distinction between conventional resonant inertial mass sensing and mass spectrometry becomes blurred.

In summary, we have demonstrated the unprecedented mass sensitivity of NEMS at the attogram scale in this work. The mass sensitivity of these first generation NEMS are dominated by the noise in the transducer circuitry, but the approach clearly offers near-term prospects for mass sensing of individual molecules.

**Table 1.** Calculated and experimental values of the mass responsivity in doubly clamped beam resonators.

| $\omega_0/2\pi$ (MHz) | $L \times w \times t$ (μm) | $M_{tot}$ (g) | $[|\Re|/(2\pi)]_{calc}$ (Hz/ag) | $[|\Re|/(2\pi)]_{expt}$ (Hz/ag) |
|---|---|---|---|---|
| 11.4 | 26.2×0.8×0.26 | 36×10$^{-12}$ | 0.22 | 0.50 |
| 32.8 | 14×0.67×0.26 | 9.9×10$^{-12}$ | 2.2 | 2.6 |
| 56 | 12×0.65×0.26 | 7.1×10$^{-12}$ | 5.2 | 5.1 |
| 72 | 100×65×0.26 | 6.0×10$^{-12}$ | 8.2 | 12 |

The authors gratefully acknowledge support for this work from DARPA MTO/MEMS under Caltech grant DABT63-98-1-0012.